\documentclass[aip,rsi,reprint,groupedaddress,twocolumn,floatfix]{revtex4-1}

\usepackage{graphicx}

\begin{document}

\title{Imaging Pulsed Laser Deposition oxide growth by in-situ Atomic Force 
Microscopy}

 \author{W.A. \surname{Wessels}}

 \author{T.R.J. \surname{Bollmann}}
 \email{t.r.j.bollmann@utwente.nl}

 \author{D. \surname{Post}}

\author{G. \surname{Koster}}

\author{G. \surname{Rijnders}}
 \affiliation{University of Twente, Inorganic Materials Science, MESA$^+$ 
Institute for Nanotechnology, P.O. Box 217, NL-7500AE Enschede, The Netherlands}

\date{\today}

\begin{abstract}
To visualize the topography of thin oxide films during growth, thereby 
enabling to study its growth behavior quasi real-time, we have designed and 
integrated an atomic force microscope (AFM) in a pulsed laser deposition (PLD) 
vacuum setup. The AFM scanner and PLD target are integrated in a single support 
frame, combined with a fast sample transfer method, such that in-situ microscopy 
can be utilized after subsequent deposition pulses. 
The in-situ microscope can be operated from room temperature (RT) up 
to 700$^\circ$C and at (process) pressures ranging from the vacuum base 
pressure of 
10$^{-6}$~mbar up to 1~mbar, typical PLD conditions for the growth of oxide 
films. 
The performance of this instrument is demonstrated by resolving unit cell 
height surface steps and surface topography under typical oxide PLD growth 
conditions. 
\end{abstract}

\maketitle

\section{Introduction}
A fascinating material class are the perovskite oxides due to their 
wealth in physical properties such as superconductivity, ferromagnetism, ferro- 
and dielectricity \cite{eason2009}. Induced by the discovery of high-$T_C$ 
superconductors, pulsed laser deposition (PLD) has become a 
popular thin film growth technique to fabricate high quality oxide materials 
\cite{Bednorz1986,Dijkkamp1987}.
The strength of depositing complex oxides with PLD comes from the fact that 
relative high oxygen pressures can be used while still having high kinetic 
energy ablated species, which strongly influences the films
properties. 
Development of the current $in-situ$ diagnostic tools such as high-pressure 
reflection high-energy electron diffraction (RHEED) \cite{rijnders1997}, surface 
x-ray diffraction (SXRD) \cite{eres2002} and the more rarely used optical 
spectroscopic ellipsometry (SE), enabled to observe $in~operando$ the oxide 
thin film growth.
Using RHEED e.g., the growth mode and in some conditions the number of grown 
unit 
cells can be deduced by measuring the step density over time 
\cite{neave1983,shitara1992,blank1999}.  
The specular rod in SXRD descibes layer filling when the diffuse scattering 
contains information about the spatial distribution of islands, whereas with SE 
the 
evolution of the electronic structure can be monitor during film 
growth \cite{blank1997,bijlsma1998}. 
Although these scattering techniques are well established tools for monitoring 
the growth of oxides $in~operando$, the reciprocal information contained can be 
hard to interpret and they do not allow to probe growth properties 
of individual thin film islands as the surface reflectivity signal 
typically probes and averages over a surface area of millimeters in 
size. Besides this, diffraction techniques typically require crystalline surface 
planes to enable observation at all.

For microscopic real-space observations a popular diagnotic tool available are 
the scanning probe microscopes (SPM), enabling monitoring the surface topology 
at the (sub)nanometer spatial resolution \cite{Lippmaa1998, Lippmaa2000, 
Iwaya2010, Shimizu2014}. As most perovskite oxides are insulators or have a 
large band gap, application of scanning tunneling microscopy (STM) is rather 
limited as it is based on a tunneling current flowing between sample and tip 
\cite{binnig1982, Sudheendra2007}.
Within the field of (PLD) oxide growth, microscopy analysis on surfaces is 
therefore typically done by post-deposition $ex-situ$ atomic force microscopy 
(AFM) or by $in-situ$ ultra high vacuum (UHV) AFM. 
However, microscopically probing the surface during deposition is however a 
prerequisite in order to broaden our understanding of thin oxide film growth. 
Real space microscopy during film growth would give complementary information 
besides reciprocal techniques about microscopic events in thin film growth such 
as diffusion processes, ripening, defect formation etc. by measuring the 
nucleation density and individual thin film island evolution over time in 
between subsequent deposition pulses \cite{rost2009}, as in PLD deposition and 
growth are separated in time.

A first design and demonstration of a conventional AFM operating at metal-oxide 
PLD conditions has been reported \cite{broekmaat2008, broekmaatthesis2008}, 
however the main drawback of typical AFM is the low sample throughput, as 
$in-operando$ monitoring requires high-speed AFM instrumentation. Conventional 
AFMs are slow due to the low bandwidth of the cantilever, AFM scanner and 
electronics and optical detection signal. Recently, a lot of progress has been 
made to increase the bandwidth of these components in different enviroments, 
which shows the potential of high speed AFM in PLD conditions 
\cite{rost2005, Kodera2006, ando2008,schitter2008, ando2012}.  

Here, we present the concept, specifications, design and performance of an 
atomic force microscope (AFM) in a pulsed laser deposition (PLD) vacuum setup. 
The setup consists of an in-situ AFM joined in an aluminum frame with an in 
geometrically position 
separated PLD position. Combined with a fast sample transfer system, this 
enables in-situ microscopy after subsequent deposition pulses. 
The AFM has been developed such that tapping mode (TM) and frequency 
modulated (FM) AFM can be applied at typical PLD conditions ranging from 
room temperature (RT) up to 700$^\circ$C and at (process) pressures ranging from 
the vacuum base pressure of 10$^{-6}$~mbar up to 1~mbar.
The performance of this instrument is demonstrated by resolving unit cell 
height steps of a SrTiO$_3$(001) surface at PLD conditions as well as 
the evolution of the surface topography of a grown BiFeO$_3$ film, a 
prototypical perovskite film.
We conclude this paper with an outlook towards future applications as well as 
limitations of the current design.


\section{Concept}


\begin{figure}
 \centering\includegraphics[width=8cm]{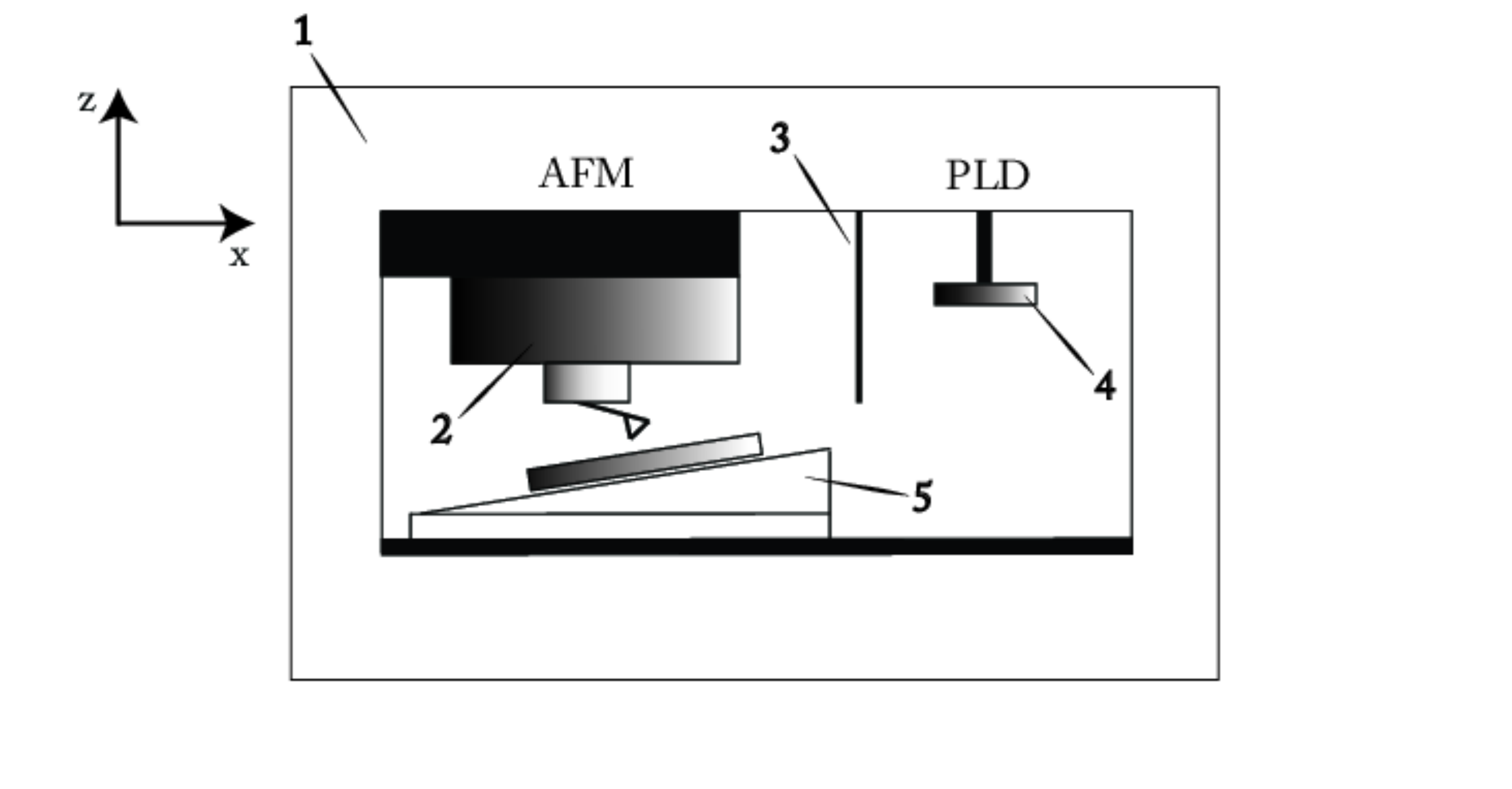}
\caption{Conceptual drawing of the PLD setup combined with $in-situ$ AFM. 
Imaging and deposition are geometrically in position seperated to prevent 
hindering of deposition and tip contamination. A fast transfer stage is moving 
the sample back and forth between imaging and deposition position. 1) support 
frame, 2) AFM scanner, 3) plasma screen, 4) PLD target and 5) sample/heater 
transfer stage.}
\label{FigAFM:concept}
\end{figure} 


Fig. \ref{FigAFM:concept} shows a schematic illustration of a concept of AFM 
during PLD. In a support frame, an AFM scanner and PLD target are installed and 
a sample transfer stage is included. The AFM scanner is positioned 
geometrically 
separated from the PLD position for a main reason: It prevents blocking of 
deposition on the as-grown surface and unwanted AFM tip contamination. As 
deposition and growth in PLD are separated in time, growth can be studied in 
between subsequent pulses using this concept. A plasma screen between AFM and 
PLD should prevent any PLD plasma contamination on the AFM scanner.

In this concept, a sample is transferred back and forth between AFM and PLD, 
where the sample transfer is based on linear motion, see 
Fig.~\ref{FigAFM:concept}. A sample heater is mounted 
on top of a linear motion slider. The concept is such that the side approach  
\cite{wessels2013} can be applied to reduce the time delay between PLD 
and AFM monitoring of the sample surface. 
The proposed configuration is straightforward to combine PLD with other 
diagnostics tools, such as scattering techniques as RHEED and plasma 
diagnostic tools for an increased understanding 
of thin film growth in PLD.

\section{Specifications}
To image the surface topography at PLD conditions, the AFM has to be operated 
at pressures ranging from 10$^{-6}$ mbar up to process pressures of 
~10$^{-1}$~mbar. The applied gases typically involve (a mixture of) 
oxygen, argon and/or nitrogen.

The temperature window of typical growth conditions runs from room temperature 
(RT) up to about ~700$^\circ$C, where the exact substrate temperature depends 
on the substrate material and ablated species. 

AFM imaging, resolving the growth properties of individual thin film islands 
requires a stable AFM imaging at an image size of about 1$\times$1$\mu$m$^2$
A short mechanical loop between the to be imaged substrate and AFM 
cantilever is required such that the noise levels, both, electronic and 
vibrational, should not exceed the substrate stepheight, typical $<$0.4~nm as 
typical oxide substrate steps are around 0.4~nm in height. This resolution is 
required both out-of-plane and in-plane.

To suppress thermal drift, an effective temperature stabilization is required. 
A thermal drift out-of-plane and in-plane, which enables measurement during 
~1~h would be required, as it is a typically duration involved in a PLD 
experiment. As 
we want to make use of image sizes of about 1$\times$1~$\mu$m$^2$, this 
requires any in-plane drift to stay below ~50~nm/min, thereby requiring minimal 
manual correction for imaging. The perpendicular thermal drift should stay 
within the piezo range to circumvent the requirement of a slow re-approach 
procedure.

To visualize the rapid growth processes involved, one needs to scan at 
high-speed, typically an image within seconds would be desirable, possibly even 
faster.

In order to reduce the dwell time between deposition and imaging, a fast sample 
transfer is required to minimize the time between geometrically seperated PLD 
position and AFM imaging. The repositioning error for subsequent back and forth 
motions has to stay below 100~nm.

The requirements can be summarized as follows:

\begin{itemize}

\item \textit{Pressures during imaging:} Ranging from 10$^{-6}$~mbar (vacuum) up 
to 10$^{-1}$~mbar in background gasses of (a mixture of) 
oxygen, argon and/or nitrogen.

\item \textit{Temperature range for imaging:} RT up to 700$^\circ$C.

\item \textit{Imaging resolution:} step resolution at oxide substrates 
($<$0.4~nm) at all mentioned pressure and temperature conditions.

\item \textit{Imaging rate:} on the order of (tens of) seconds per image of 
256$\times$256 pixels with images of about 1$\times$1$\mu$m$^2$.

\item \textit{Thermal drift:} below 700~nm/h in perpendicular direction, below 
50~nm/min in-plane.

\item \textit{Transfer time before and after imaging:} transfer time back and 
forth within 0.5~s with a repositioning error below 100~nm.

\end{itemize}

\section{Design}

In this section, the general architecture of the vacuum setup, the designed 
\emph{in-situ} AFM within a PLD vacuum chamber and the design of the AFM 
scanhead is described in detail. 

\subsection{Vacuum chamber and peripheral equipment}


\begin{figure}
\centering
\includegraphics[width=8cm]{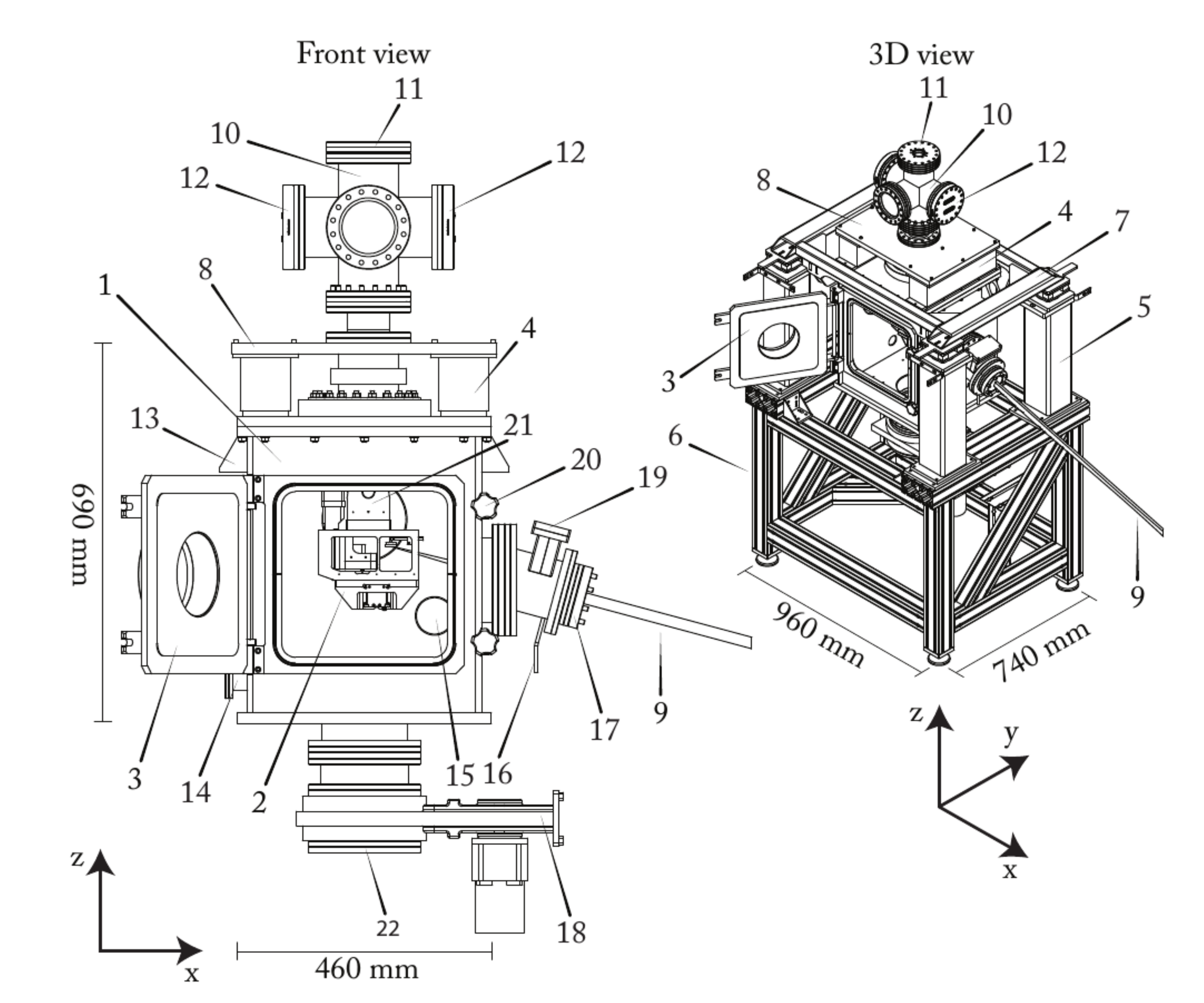}
\caption{Schematic drawing of the vacuum setup and peripheral equipment. 1) 
PLD vacuum chamber, 2) AFM-PLD frame, 3) door of chamber, 4) Halcyonics 
MOD2-S active damping control element, 5) air damped pods, 6) Aluminum profile 
system, 7) passive damping stage, 8) active damping stage, 9) incoming 
beampath for the excimer laser, 10) DN100CF five-way cross, 11) fiber 
feedtrough, 12) electrical feedtrough, 13) mounting of the passive damping 
stage, 14) and 15) DN40CF flanges for pressure sensors,
16) gasinlet, 17) DN63CF flange with quartz glass, 
18) adaptive pressure controlled valve, 
19) valve for laser energy measurement, 
20) closing unit door, 
21) Aluminum tube for feedtrough and connecting the active damping stage with 
the supporting frame, 
22) DN200CF flange towards turbopump. 
}
\label{FigAFM:vacuumchamber}
\end{figure}


Fig. \ref{FigAFM:vacuumchamber} shows a schematic representation of the 
designed PLD chamber and peripheral equipment. The PLD chamber is a 
single vacuum system, which differs in shape compared to traditional 
cylindrical chambers. In the designed vacuum system, the base pressure of 
10$^{-7}$~mbar is limited by O-rings of the vacuumdoor, valve 
on the flange on the laser beam path and the pressure controled valve. Typical 
PLD background pressures of 10$^{-6}$~-~1~mbar are measured by almost 
closing this valve and varying the incoming gas flow between 
zero and 100~ml/min. The vacuumchamber is a square-like box design such that 
a vacuum chamber door is used to open up the system for accessibility to 
install the supporting frame for AFM and deposition. The vacuumdoor 
has a DN100CF flange viewport window, revolves by a shaft and can be closed by 
two closing knobs. Four airpod damping units together with matted corks placed 
underneath provide passive damping of system resonances validated by measuring 
the transfer function in x- and 
y-direction \cite{segerink2011}. 
The vacuum chamber houses an aluminum tube connecting the active damped system 
towards the supporting frame for AFM and deposition. The active damping stage 
is mounted as a lid on this PLD chamber ensuring good 
vibration isolation.

On top of the active damping stage, a five-way cross has been assembled 
providing feedtroughs for optical and electrical signals from and to the 
AFM scanner, coarse approach steppermotor, slider piezomotor drive and heater. 
The DN40CF flange ports are used for vacuum pressure gauges to measure both, 
process pressures as well as (high) vacuum pressures. 

A KrF excimer laser beam aligned to an optical rail is integrated such that 
a 248~nm pulsed laser beam, having a typical pulse duration of 25~ns, is 
focused on the PLD target. As an entrance flange for the PLD excimer laser 
light, a flange is designed to enable the alignment of the laser 
onto the PLD target in an angle of 15$^\circ$ parallel to the targets surface. 
The flange is integrated with a valve to support 
laser intensity measurements behind the quartz window as well as a gasinlet for 
the applied background gasses. A flow meter controlled gas manifold has been 
designed to separately let (a mixture) of nitrogen, oxygen and/or argon into 
the 
vacuum chamber through this gas-inlet. 
At the bottom of the vacuum chamber, a DN200CF flange is connected 
to an adjustable valve typically used for automated pressure control followed 
by a turbomolecular pump. 

\subsection{Support frame for microscopy and deposition}

\begin{figure}
 \centering
 \includegraphics[width=8cm]{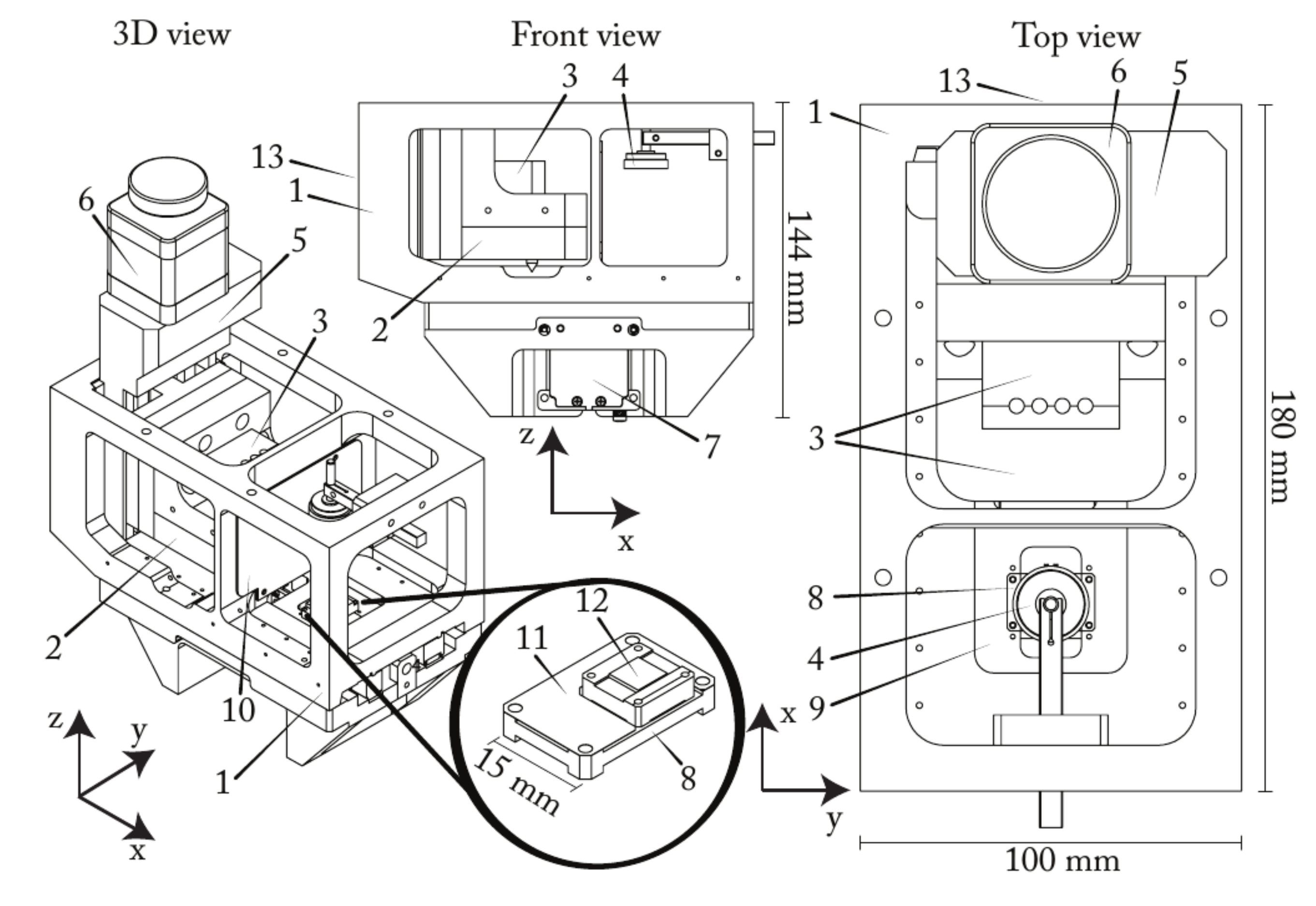}
\caption{Schematic drawing of support frame including AFM and PLD. 1) Al frame, 
2) AFM flexure scanner, 3) mounting bracket, 4) PLD target 5) coarse approach 
stage, 6) Vexta 5 phase stepper motor PK545-B, 7) HR4 nanomotion ultrasonic 
piezomotor, 8) macor heater stage, 9) sample transfer stage, 10) plasma screen, 
11) two Kamet HDA pt200 RTD's connected in serie, 12) 5$\times$5~mm$^2$ sample 
on top of an 5$\times$5~mm$^2$ Pt plate and 13) location of screwed cube for 
integration of acceleration level sensors.}
\label{FigAFM:supportframe}
\end{figure}

Fig. \ref{FigAFM:supportframe}, schematically illustrates the supporting frame 
which contains the AFM and deposition stage. This aluminum frame, exhibiting a 
robust and high stiffness design, is attached by an aluminum tube to an active 
damping stage, see also Fig.~\ref{FigAFM:vacuumchamber}. 
This damping stage has the functionality to lower the vibrational level 
in the mechanical loop of the AFM. 
The chosen shape and size of the support frame is based on required sizes and 
shape of the AFM scanner, PLD deposition and highest mechanical stiffness such 
that the required vibrational level is not exceeded. 
The aluminum frame consist of two assembled frames, where the top frame is 
used for integration of both the AFM scanner and PLD target. 
The $z$-approach stage, containing a stepper motor having a step resolution 
of 100~nm/step, is positioned in the top frame to provide coarse approach 
towards the sample. A homebuilt mounting bracket connects the coarse approach 
stage to the AFM scanhead. The PLD target can be mounted onto a 
holder, adjustable in both height and lateral position along the sample 
transfer direction. 
In this design, the distance between the AFM tip and center of the PLD 
target is typically 50-60~mm, where the PLD target is shielded by a 
plasmascreen 
from the AFM scanhead.

The bottom frame has the functionality to integrate a sample transfer stage 
enabling back and forth sample transfer between the geometrically seperated AFM 
and deposition stage. 
A vacuum piezomotor with low vibrational level, a repositioning resolution 
within 20~nm and a maximum speed of 244~mm/s is installed in the bottom frame 
in order to propel the sample transfer stage which is designed to have minimal 
mass. 


\subsection{AFM flexure scanner}

\begin{figure}
 \centering
 \includegraphics[width=8cm]{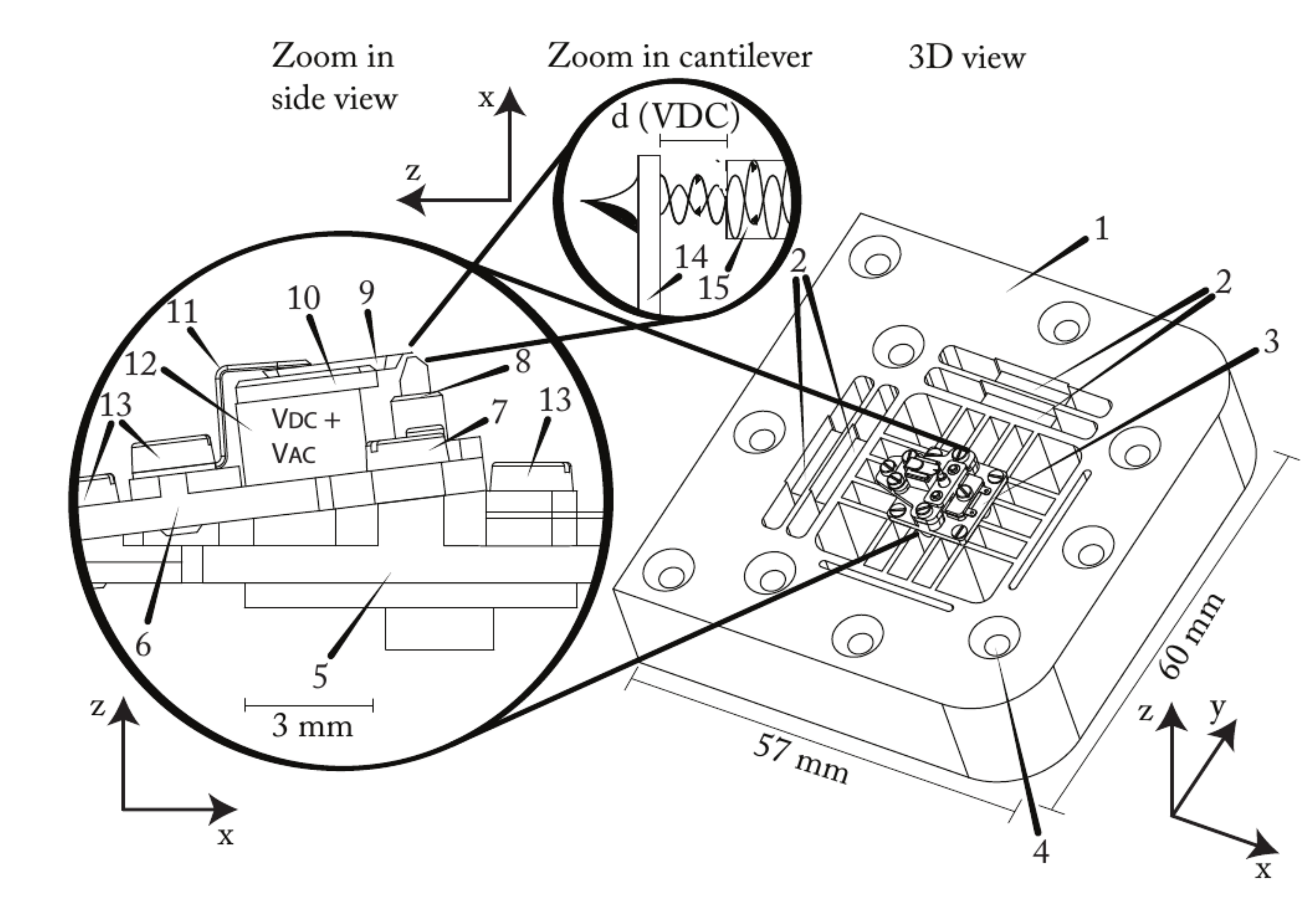}
\caption{Schematic drawing of the high resonance frequency AFM flexure scanner, 
with 
1) the Ti6Al4V flexure, 
2) stacked XY piezo actuators, 
3) AFM chip holder, 
4) screw holes, 
The insets shows the:
5) connection plate between AFM chip base plate and the flexure, 
6) cantilever base plate
7) electrical contacts of the dither piezo, 
8) ferrule, 
9) cantilever chip, 
10) chip holder, 
11) cantilever clamping spring, 
12) dither piezo stack, 
13) screws, 
14) AFM cantilever, 
15) optical fiber end. 
}
\label{FigAFM:AFMscanner}
\end{figure}

The AFM scanner is based on a flexure, see Fig. \ref{FigAFM:AFMscanner}, which 
is often used in the field of high speed AFM~\cite{ando2012}. Flexure scanners 
are used as an alternative to commonly used low resonance frequency piezotubes 
to extend the mechanical scanner bandwidth several orders of magnitude into
the kHz range~\cite{yong2012}. 
In addition, stacked piezo's integrated in a flexure results in less 
cross-coupling, hysteresis and creep as compared to generally used piezotubes. 
The AFM flexure in this work has been made out of Ti6Al4V to achieve high 
stiffness, a resulting high resonance frequency and low thermal expansion in 
order to minimize thermal drift. A simulation using finite element 
analysis resulted in a lowest resonance of 19~kHz for the flexure in 
the XY-plane.
The maximum achievable linerate, estimated as 1/100th~-~1/10th of the lowest 
resonance frequency \cite{yong2012}, is therefore between 0.19~-~1.9~kHz for 
this scanner design.  

In order to achieve the desired scan range, we make use of piezo actuators with 
an expansion up to 4~$\mu$m, which strongly reduces up to 1.4~$\mu$m due to the 
high preload. The Z-flexure, a piezoring with an effective scan~range of 
1.4~$\mu$m is aligned nearly in the center of the XY-flexure and can be 
positioned with high precision within the XY-plane by the double stacked 
XY-actuators.
A counterbalance piezo, integrated in the Z-flexure, counters the momentum 
generated by center of mass movement of the Z-piezo. 
The unit, holding the AFM chip, is displaced by this Z-piezo such that the 
actual cantilever is excited, see the inset in Fig.~\ref{FigAFM:AFMscanner}.

The bottom part of this unit is screwed on top of the Z-flexure and is milled 
under an angle of 7$^\circ$. On top of this a flat plate with dither piezo 
stack is mounted. To detect the cantilevers deflection we make use of optical 
interferometry through the ZrO$_2$ ferrule shown in 
Fig.~\ref{FigAFM:AFMscanner} instead of the more common optical beam deflection 
method.
Beam deflection requires use and adjustment of optical instrumentation, such as 
mirrors and is unpractical under PLD conditions, where space is lacking. 
Therefore, usage of interferometric detection results in a more compact design. 
A laser beam is reflected from the cantilever top side and fiber-air interface, 
where the cantilever and fiber are separated by 40-50~nm. A reference 
wave is reflected by the planar end of the fiber (glass-air interface). The 
detected wave is reflected by the top face of the cantilever. Both light waves 
interfere on a photodiode for detection.
The optimal interference working point distance of the cavity can be adjusted 
by a DC voltage over the dither piezo. On top of this $V_{\mathrm{DC}}$, an 
AC voltage (typically $V_{\mathrm{AC}}$~=~1~mV~-~1~V) is applied to 
oscillate 
the cantilever at its resonance frequency $f_{0}$ in dynamic mode AFM.

\section{Performance}
\label{sec:performancesetup}
In this section we demonstrate the performance of the crucial components of the 
experimental setup. We focus here on the vibrational level and characteristics 
of the setup under PLD conditions. We then turn to the imaging performance by 
demonstrating the imaging of a typical oxide surface, SrTiO$_{3}$(001), under 
some typical conditions experienced in a PLD experiment as well as the imaging 
of the growth of a BiFeO$_{3}$ film under its PLD growth conditions. 

\subsection{Vibrational level and stage translation}

\begin{figure}
 \centering
\includegraphics[width=8cm]{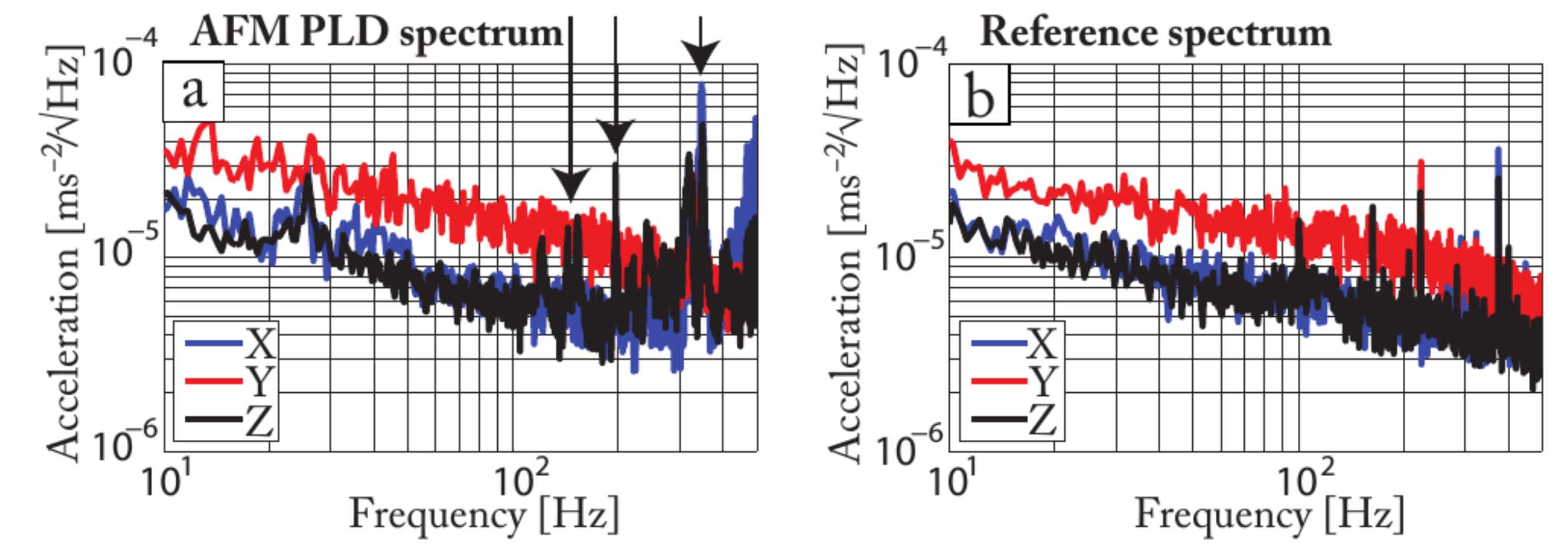}
\caption{Vibrational spectra of the, (a) designed setup at a typical oxygen PLD 
pressure of 10$^{-1}$~mbar~O$_{2}$ in comparison to a commercial (b) AFM setup. 
The peaks marked by arrow correspond to resonance frequencies of the aluminum 
support frame.}
\label{FigAFM:accelerationlevel}
\end{figure}

Fig.~\ref{FigAFM:accelerationlevel} shows the X,Y,Z~acceleration levels of the 
AFM-PLD setup operating at 0.1~mbar oxygen background pressure in comparison 
to that of the acceleration levels of a commercial AFM setup (Bruker Dimension 
Icon) operating in ambient conditions. 
Both systems are placed on the same VC-G classified floor. The vibrational level 
sensors are positioned in the mechanical loop from tip to the sample, see 
Fig.~\ref{FigAFM:supportframe}. The passive damping system suppresses the higher 
frequencies $\geq$~100~Hz, whereas the active damping stage significantly 
decreases acceleration levels from 10~-~100~Hz. Damping of the system 
by the airpods is mainly done in the Z-direction. There is no significant 
influence in acceleration levels observed for the peripheral equipment. 
Acceleration levels of the designed setup are similar to the commercial AFM 
setup, see Fig.~\ref{FigAFM:accelerationlevel}. Only resonance frequencies 
of the aluminum support frame, see Fig.~\ref{FigAFM:vacuumchamber}, are found 
at 140,~190 and several peaks between 300~-~400~Hz, marked by arrows in 
Fig.~\ref{FigAFM:accelerationlevel}.

Crucial for the designed setup, is the performance (directly) after stage 
transfer. We therefore recorded the time domain signal together with the 
acceleration levels during such transfer. 
The vibrational decay time of the mechanical AFM tip-sample loop is determined 
from the time domain signal from the moment that the transfer stage velocity is 
zero at the AFM position after sample transfer from PLD position to the AFM 
position, for more details we refer to our earlier report \cite{wessels2013}. 
In the parallel acceleration level measurements, resonance peaks are observed at 
140 and 280~Hz directly after stage deceleration and a corresponding delay 
time of 0.4~s. The back and forth sample transfer between AFM and PLD is 
possible in 0.5~s driving at maximum acceleration and velocity. 


\subsection{The role of PLD conditions}
\label{subsec:AFM-PLD}
Typical PLD conditions involve substrates at elevated temperatures and process 
pressures up to several mbar, a pressure regime where cantilever response is 
known to alter dramatically \cite{Lubbe2011}.
The cantilever resonance frequency $f_{0}$ and the cantilever quality 
$Q$ factor are nearly constant at pressures ranging from 
10$^{-6}$~-~0.1~mbar \cite{mertens2003}. However, $Q$ 
increases from a few hundred to tens of thousands in the pressure regime 
1~bar~-~0.1~mbar \cite{Lubbe2011}, corresponding to a broadening in its 
frequency 
response.

Also the elevated substrate temperatures involved in PLD growth, can result in 
temperature variations of the cantilever during stage transfer. These 
temperature variations are known to result in unwanted drift of the eigen 
frequency $f_0$ of the cantilever as the cantilevers dimensions alter and the 
Young's modulus of silicon is known to vary reasonably upon temperature. 
Measuring $f_0$ 
 at RT and within close vicinity of a sample heated up to 
600$^\circ$C at a process pressure of 10$^{-1}$~mbar resulted in a 
frequency shift $\Delta f$ of 200~Hz, corresponding to a temperature increase 
of the cantilever of $\approx$~70~$^\circ$C. By use of cantilevers with 
reasonable but rather low $Q$ factors, the used side-approach 
\cite{broekmaat2008} upon stage transfer can still be made succesfull, having 
the compromise of lower resolution. By the use of a radiation shield we 
prevented thermal radiation to influence the working point distance of the 
interferometric setup

\subsection{AFM imaging}
\begin{figure}
\centering\includegraphics[width=8cm]{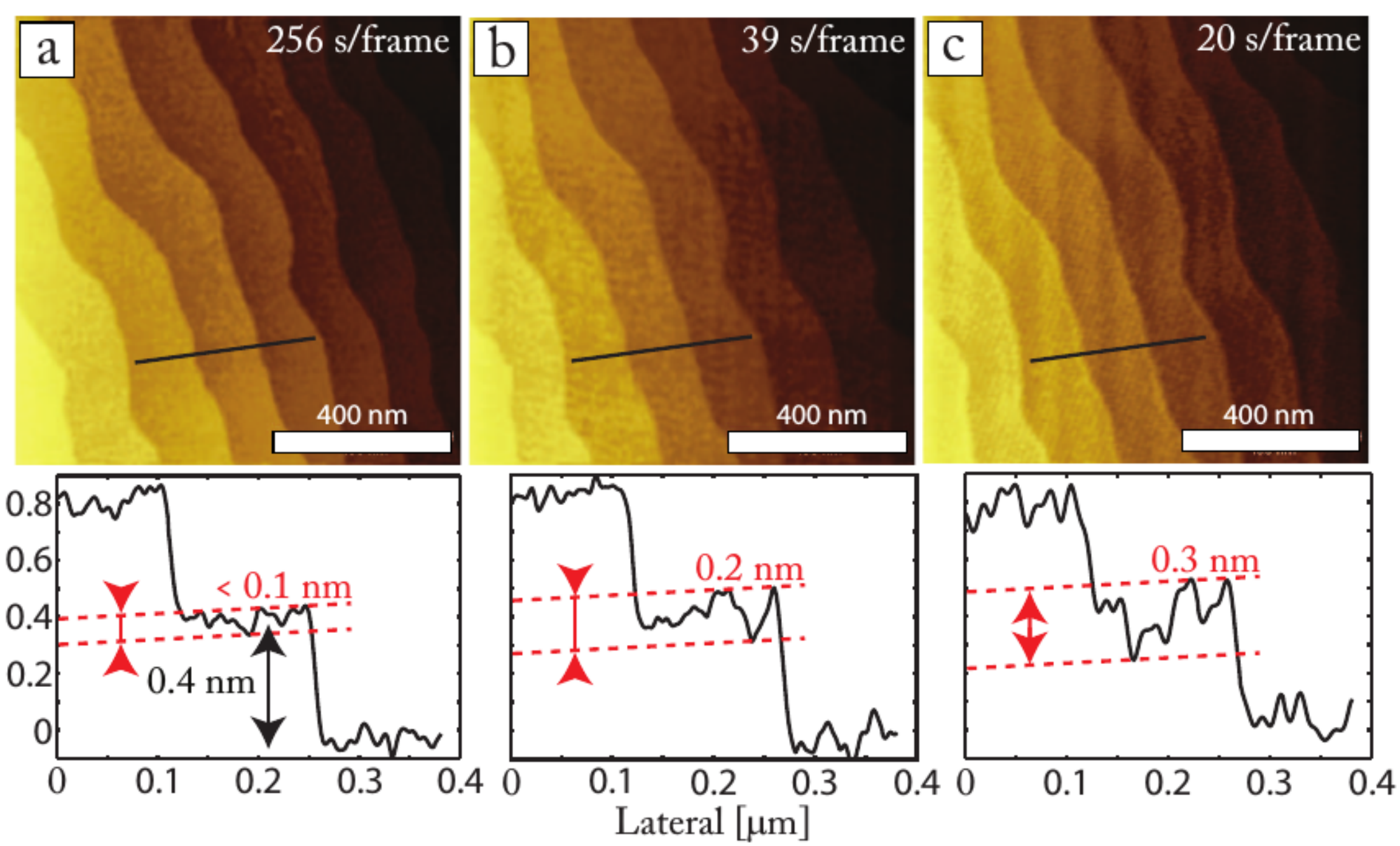}
\caption{(Color online) TM-AFM 1$\times$1~$\mu$m$^2$ 512$\times$512 pixels$^2$
images of a SrTiO$_3$(001) substrate in air at RT with in the bottom panels the 
corresponding height profiles. The scan speed is (a) 256~s/frame, (b) 
39~s/frame and (c) 20~s/frame. The resolution decreases significantly at a scan 
speed of 20~s/frame due to the limitation of the used cantilever resonance 
frequency $f_0$.
}
\label{FigAFM:AFMspeed}
\end{figure}

In order to demonstrate the scan speed performance we performed tapping 
mode (TM)-AFM measurements in air, shown in Fig.~\ref{FigAFM:AFMspeed}. The 
scan speed is increased for figures Fig.~\ref{FigAFM:AFMspeed} (a-c) measuring 
the SrTiO$_3$(001) substrate steps of 0.4~nm in height. At conventional AFM 
scan speeds, having a 2~Hz line rate (corresponding to an acquisition 
time of 256~s/frame), the subnanometer SrTiO$_3$ substrate steps are clearly 
resolved, see also the height profiles in Fig.~\ref{FigAFM:AFMspeed}. From this 
line-profile, a peak-to-peak RMS of $<$0.1~nm was found. Upon reducing the 
acquisition time, see Fig.~\ref{FigAFM:AFMspeed}(b) and (c), towards 
respectively 39~s/frame and 20~s/frame, this 
peak-to-peak RMS increases towards $\approx$0.3~nm.
In addition, a resonance frequency of 1.6~kHz is appearing, a vibrational mode 
of the mounting bracket connection with the coarse approach stage.

The increase in peak-to-peak z-noise can be attributed to two sources; 
firstly the electronic noise which increases proportional to $\sqrt{f_{bw}}$, 
note that the used electronics are known to deliver sufficient resolution at 
video 
rate and beyond \cite{rost2005}. Secondly, the currently used photodetector 
which is limited to 400~kHz, thereby practically limiting to cantilevers of 
$\sim$~300~kHz. 
A reduced pixel dwell time results in less 
oscillations used to determine a RMS value for every single pixel.

\begin{figure}
\centering\includegraphics[width=8cm]{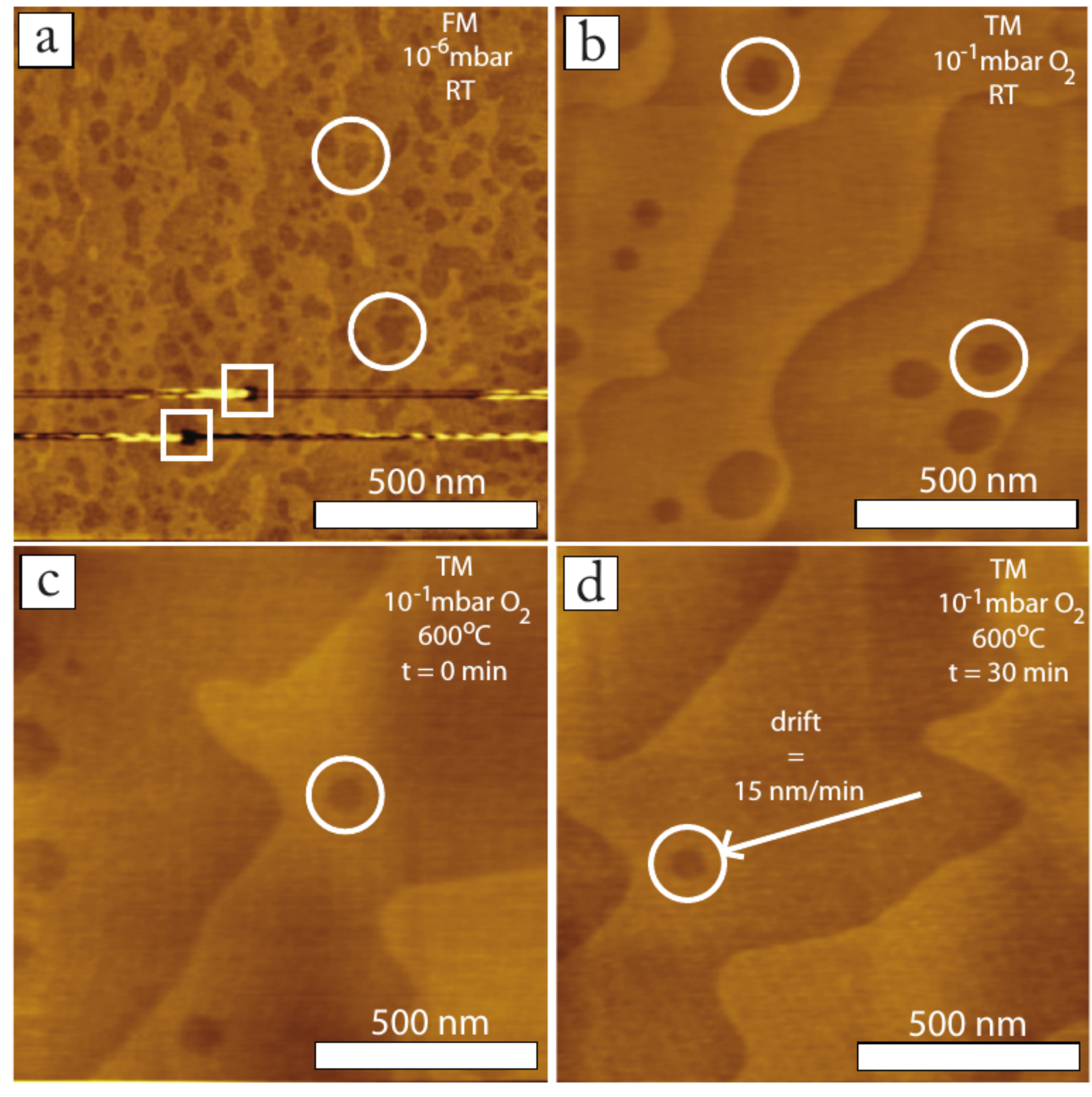}
\caption{(Color online) In situ AFM 1.2$\times$1.2 $\mu$m2 (512*512 
pixels$^2$) images of SrTiO3(001) substrates under different conditions (and 
different miscut angles) recorded with an acquisition time of 256 s/frame. a) 
FM-AFM image, $\Delta$f = -28 Hz, $A_{osc}$ = 10 nm, at RT and P = 10$^{-6}$ 
mbar, b) TM-AFM image, A$_{osc}$ = 44 nm, at RT and P$_{O2}$ = 10$^{-1}$ mbar, 
c) TM-AFM image, Aosc = 44 nm, at T = 600$^\circ$C and P$_{O2}$ = 10$^{-1}$ 
mbar O$_2$, d) TM-AFM image, A$_{osc}$ = 44 nm, at T = 600$^\circ$C and 
P$_{O2}$ = 10$^{-1}$ mbar O$_2$ after 30 min AFM imaging. Circle markers 
surround unit cell vacancy islands with a depth of 0.4 nm and square markers 
surround etch pits of several unit cells deep.
}
\label{FigAFM:AFMmeasurements}
\end{figure}

Typical PLD conditions involve process pressures ranging from high vacuum 
conditions up to 1~mbar. To demonstrate the use of the microscope at these 
conditions, in Fig.~\ref{FigAFM:AFMmeasurements} we show AFM images recorded 
under typical PLD conditions. In Fig.~\ref{FigAFM:AFMmeasurements}(a), a 
frequency-modulated (FM) AFM image is depicted of a SrTiO$_3$(001) substrate 
containing unit cell vacancy islands of 0.4 nm deep recorded in a background 
pressure of 10$^{-6}$ mbar at RT \cite{koster1998}. In this pressure regime, 
the FM-AFM imaging mode is typically used due to the high $Q$ for image 
stability reasons and lower transient time compared to TM-AFM 
\cite{garcia2002,albrecht1991}. 
In this FM-AFM image, the vertical resolution is similar to the z-noise level 
($<$0.1~nm) and unit cell (0.4~nm) deep vacancy islands are resolved having 
lateral sizes of $\approx$20 nm. 

In Fig.~\ref{FigAFM:AFMmeasurements}(b), a TM-AFM image is depicted of a 
SrTiO$_3$(001) substrate at 10$^{-1}$~mbar oxygen background pressure , which 
is a typical 
PLD pressure to deposit perovskite oxide films ensuring good crystal quality, 
properties, stoichiometric transfer and 2D growth \cite{Groenen2015}. 
At 10$^{-1}$~mbar oxygen background pressure, the free cantilever amplitude 
 was set to $\sim$~120~nm and is found to be the most critical parameter for 
AFM 
imaging. The lateral AFM resolution on terraces is lowered as a SrTiO$_3$(001) 
step broadening of $\approx$20~nm is found at this pressure. An 
average in-plane drift of 2~nm/min is measured over a period of 3.5~hours under 
these conditions.

The same AFM settings have been used for a sample measured at T=600$^\circ$C. 
At pressures of 10$^{-1}$~mbar oxygen and sample temperatures of 
T=600$^\circ$C vacancy islands can easily be resolved. 
The AFM noise increase due to temperature is negligible based on the fact that 
z-noise level and SrTiO$_3$(001) step broadening remained at T=600$^\circ$C 
\cite{broekmaatthesis2008}. 
Under these conditions, a drift of 15~nm/min is measured as depicted in 
Fig.~\ref{FigAFM:AFMmeasurements}. After hours of stable AFM imaging under oxide 
PLD conditions, the z-piezo stayed within range ($\pm$700~nm).

\subsection{AFM imaging during and after PLD}

\begin{figure}
\centering\includegraphics[width=8cm]{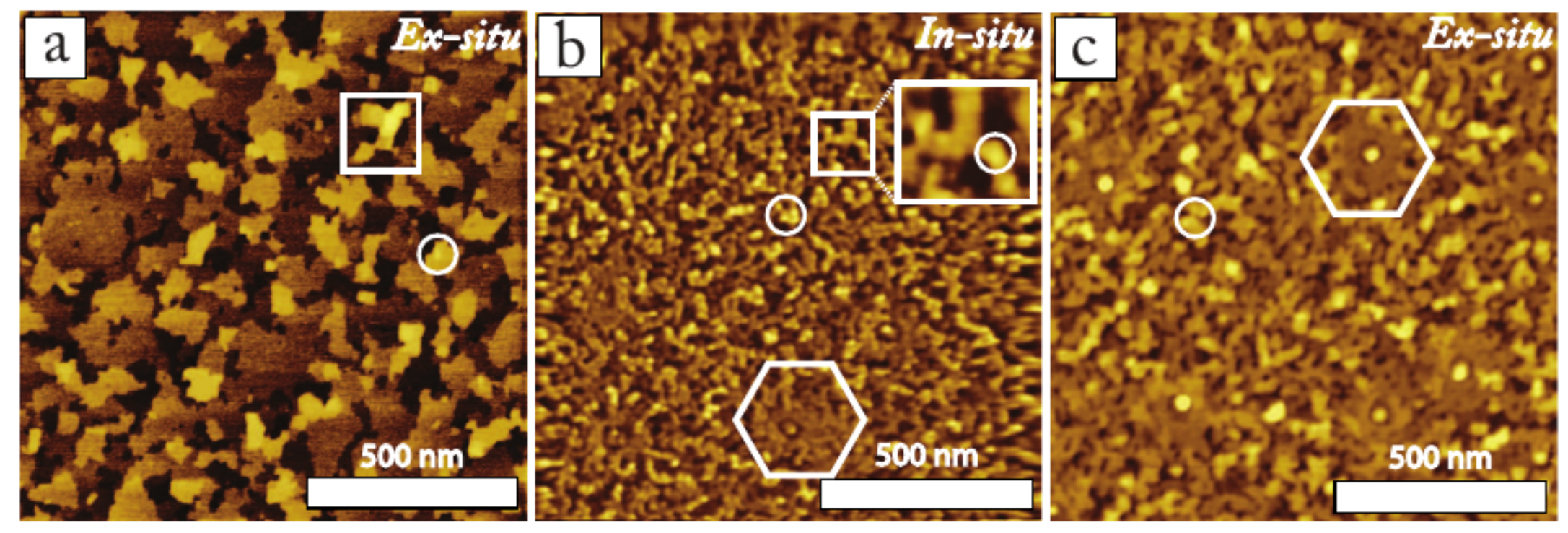}
\caption{(Color online) Fig. 3.13 TM-AFM 1.2$\times$1.2~$\mu$m$^2$ 
(512$\times$512 pixels$^2$) images, acquisition time=256~s/frame, after 1000 
pulses of BiFeO$_3$ deposited on SrTiO$_3$(001) (a) ex-situ at RT and in air 
after deposition at T=670$^\circ$C, (b) in-situ at T=600$^\circ$C and 
P$_{O2}$=0.3~mbar O$_2$ after 1000 pulses of BiFeO$_3$ deposited on 
SrTiO$_3$(001), (c) ex-situ at RT and in air after deposition at 
T=600$^\circ$C. Square markers point to 3D BiFeO$_3$ islands, circle markers 
point to small BiFeO$_3$ islands and polygon markers point to larger islands.}
\label{FigAFM:AFMPLD}
\end{figure}

BiFeO$_3$ has been deposited as a prototypical ABO$_3$ film using the described 
PLD system. This material has multiferroic properties and its growth and 
property relation is intensively studied 
\cite{lim2008,bea2011,kanashima2014,deepak2015,solmaz2016}. It is 
reported that BiFeO$_3$ tends to grow in several types of domains by modifying 
the SrTiO$_3$(001) termination 
\cite{solmaz2016}. Dependent on the substrate termination, BiFeO$_3$ grows 
either 1D/2D 
or 3D. Here, BiFeO$_3$ films have been grown on TiO$_2$ terminated 
SrTiO$_3$(001) substrates anticipating 3D growth. BiFeO$_3$ was deposited with 
a 
laser fluence of 2.0~J/cm$^2$ , 0.3~mbar oxygen background pressure, a sample 
temperature of 600-670$^\circ$C during deposition and a 
repetition rate $f_{rep}$=0.5~Hz, all settings similar to previous work where 
only a target-substrate distance of 45 mm is used instead of 55 mm 
\cite{solmaz2016}.
Fig.~\ref{FigAFM:AFMPLD} shows AFM images obtained after and during BiFeO$_3$ 
deposition. In Fig.~\ref{FigAFM:AFMPLD}(a), an ex-situ AFM image is depicted 
of a BiFeO$_3$ film after deposition of 1000 pulses at 670$^\circ$C with a film 
thickness of $\approx$3-4~nm. This BiFeO$_3$ film was deposited as a reference 
to reported literature \cite{solmaz2016}. It is reported that BiFeO$_3$ grows 
3D 
on 
TiO$_2$ terminated SrTiO$_3$(001), which is similar to the results obtained 
here. Terraces with on top 3D islands are visible after 1000 PLD pulses of 
BiFeO$_3$ deposited on TiO$_2$ terminated SrTiO$_3$(001). The 3D BiFeO$_3$
islands on top of the terraces suggest that next layer nucleation starts before 
a previous layer is completely covered. Island step heights are found of 
0.2~nm, 0.4~nm and its multiples up to a maximum peak-to-peak height of 3~nm. 
Islands of 25-30~nm in lateral size are obtained on top of random shaped 
islands. From these results, it is expected that BiFeO$_3$ growth will continue 
3D, similar to what has been reported \cite{solmaz2016}. 

In 
Fig.~\ref{FigAFM:AFMPLD}(b), an in-situ AFM image is presented, which was 
recorded at T=600$^\circ$C, as this is the maximum stable operating temperature 
for AFM measurements.
The process pressure used was 0.3~mbar oxygen background pressure depositing 
again 1000 pulses of BiFeO$_3$. The AFM image was taken after BiFeO$_3$ 
deposition, sample 
transfer from PLD to AFM and an AFM stabilization time of 2h at the AFM 
position. After thermal stabilization, AFM was started with similar settings 
used for imaging of a SrTiO$_3$(001) substrate. It became clear that due to an 
increase in surface roughness after deposition, the integral gain of AFM 
electronic feedback had to be increased significantly to visualize the smallest 
BiFeO$_3$ islands. Stable AFM imaging was continued for several hours on 
BiFeO$_3$ without thermally drifting out of the z-range. Thermal drift causes 
small distortions at AFM image edges. Smallest lateral BiFeO$_3$ island
sizes of 20-30~nm have been measured, similar to the results obtained with an 
ex-situ AFM on BiFeO$_3$ after deposition at T=670$^\circ$C \cite{solmaz2016}. 
Some 
of these small islands are surrounded by a white circle marker in 
Fig.~\ref{FigAFM:AFMPLD}. 
In Fig.~\ref{FigAFM:AFMPLD}(b), the square white marker represents a zoom-in 
of a BiFeO$_3$ taken from an 0.8$\times$0.8~$\mu$m$^2$ AFM image.

Afterwards, the BiFeO$_3$ film grown on SrTiO$_3$(001) at T=600$^\circ$C had a 
cooldown of $\approx$15$^\circ$C/min in its deposition pressure of 
0.3~mbar oxygen background pressure. Once the sample reached RT, it was exposed 
to a maximum oxygen flow up to atmospheric pressure.
In Fig.~\ref{FigAFM:AFMPLD}(c), an ex-situ AFM image is depicted of BiFeO$_3$ 
after oxygen exposure up to atmospheric pressure. The AFM image in 
Fig.~\ref{FigAFM:AFMPLD}(c) is slightly different
compared to Fig.~\ref{FigAFM:AFMPLD}(b). One difference is that BiFeO$_3$ 
islands of $\approx$20-30~nm are hardly visible in the ex-situ AFM image, see 
Fig.~\ref{FigAFM:AFMPLD}(c). Both AFM 
images have in common that BiFeO$_3$ islands within polygon markers are similar 
in size. 
Note that, $Q$ decreases more 
than an order of magnitude from 10$^{-1}$~mbar background pressure up to 
atmospheric 
pressure. 

This study reveals that BiFeO$_3$ can be imaged using a Si AFM tip at 
T=600$^\circ$C and 0.3~mbar process pressure after deposition on a TiO$_2$
terminated SrTiO$_3$(001) substrate. Neck formation has not been observed using 
this tip-sample combination under these conditions. One of the ideas is to 
deposit BiFeO$_3$ on mixed terminated SrTiO$_3$(001) in order to study growth 
(front evolution) differences on both SrO and TiO$_2$ termination. However, AFM 
imaging (Si tip) on mixed terminated SrTiO$_3$ substrates results in neck 
formation at T=600$^\circ$C and 0.1~ mbar process pressure, while stable AFM 
imaging was achieved on a TiO$_2$ deposited film on TiO$_2$ terminated
SrTiO$_3$(001) under the mentioned conditions. From the described measurements
and the neck formation of a Si AFM tip with a SrRuO$_3$ film (SrO termination) 
at T=600$^\circ$C and 0.1~mbar oxygen background pressure, it seems plausible 
to argue that 
a Si AFM tip forms a neck if the surface contains a SrO top layer 
\cite{rijnders2004,broekmaatthesis2008}. 
For this tip-sample combination, another tip material/tip coating needs to be 
selected.

\section{Conclusions and outlook}
To visualize the topography of thin oxide films during growth, we have 
designed and integrated an atomic force microscope (AFM) in a pulsed laser 
deposition (PLD) vacuum setup. The in-situ microscope is demonstrated to 
operate at typical PLD conditions, thereby resolving unit cell height surface 
steps and surface topography.

We end this paper by discussing some aspects that might improve the 
performance of this microscope. The setup described here is a major step 
towards real-time AFM during PLD conditions. To enable quasi real-time AFM 
monitoring of island growth during PLD, the bandwidth of both the cantilever and 
(optical) detection system need to be improved, as the acquisition rate is 
currently limited by them. As high resonance cantilevers dictate smaller 
physical dimensions on the cantilever itself, the optical detection system might 
also need reconsideration. We are in the process of the development of 
self-sensing piezo-electric cantilevers in order to increase the bandwidth by 
(an) order(s) of magnitude. 

Besides this, the thermal drift of the cantilever resulting from varying 
temperature gradients 
during sample transfer also needs reconsideration as slight temperature 
variations of the cantilever might result in a considerable eigenfrequency 
shift 
of it. Moreover, increasing both the AFM feedback bandwidth and transfer stage 
speed will result in a faster tip-sample approach time using the discussed side 
approach. We therefore currently develop a modification of the geometry such 
that the temperature of both sample and cantilever is better controlled 
together with a faster approach.

\section{Acknowlegdement}

This work is part of the research programme of NanoNext NL project 9A 
nanoinspectation and characterization, project 07 real-time atomic force 
microscopy growth monitoring during pulsed laser deposition of oxides. 

\section{References}

%


\end{document}